\documentclass[aps,prl,twocolumn,]{revtex4}
\usepackage{amssymb}
\usepackage{amsmath, amsthm}

\begin{document}


\title{Solar System experiments do not yet veto modified  gravity 
models}


\author{Valerio Faraoni}
\email[]{vfaraoni@cs-linux.ubishops.ca}
\affiliation{Physics Department, Bishop's University\\
Sherbrooke, Qu\`ebec, Canada J1M~0C8
}


\date{\today}

\begin{abstract}
The dynamical equivalence between modified and 
scalar-tensor gravity theories is revisited and it is concluded 
that it breaks down in the limit to general relativity. A 
gauge-independent analysis of cosmological perturbations in both 
classes of  theories lends independent support to this 
conclusion. As a  consequence, the PPN  formalism of 
scalar-tensor gravity and Solar System  experiments do not 
veto modified gravity, as previously thought.
\end{abstract}


\maketitle

\section{Introduction}
\setcounter{equation}{0}
There are many models in the literature aiming at explaining the 
observed acceleration of the cosmic expansion discovered with 
supernovae of type Ia \cite{SN} in conjunction with the 
recent cosmic microwave background experiments \cite{CMB}. One 
class of models 
postulates 
that the universe is filled with unclustered dark energy 
comprising 70 percent of its energy content. This dark energy has 
exotic properties and, if its effective equation of state is 
truly  such that $P<-\rho$ (where $\rho$ and  $P$ are the dark 
energy 
density 
and pressure, respectively), as the observations seem to favour 
\cite{w}, 
it is even more exotic and it is called phantom energy or 
superquintessence. Phantom energy violates all of the energy 
conditions and is rather difficult to accept because of the 
possibility of instabilities, ghosts, and strange thermodynamical 
behaviour \cite{phantom}. As an alternative to 
such exotic physics, it has been 
suggested that perhaps gravity should be modified at large scales 
\cite{MG1,CDTT} by introducing in the gravitational sector terms 
non-linear in the 
Ricci curvature $R$. This way, one 
can dispense entirely with exotic dark energy. Apart 
from this {\em ad hoc} justification, there are also 
motivations (and corrections) for non-linear gravity from 
M-theory \cite{Mtheory}. Scenarios based 
on this idea are called ``modified gravity'', ``non-linear 
theories'', or ``fourth-order gravity'' \cite{MG3}. In its 
simplest form, 
the action is
\begin{equation} \label{1}
S=\frac{1}{2\kappa}\int d^4x \, \sqrt{-g} \, f(R)+S^{(m)}
\end{equation}
where $\kappa \equiv 8\pi G$. The corresponding field equations  
are 
\begin{equation} \label{1bis}
f' R_{ab}-\frac{f}{2}\, g_{ab}=\nabla_a\nabla_b f'-g_{ab} \, \Box 
f'+\kappa T_{ab}^{(m)} \;,
\end{equation}
where  a prime denotes differentiation with respect to $R$.  
Whenever $f(R)$ is non-linear in $R$, the Palatini variation 
treating the metric and the connection as independent 
variables produces field equations that are different  from 
(\ref{1bis}), which are  
obtained with the usual Einstein-Hilbert variation with respect 
to the metric only. The ``Palatini 
approach'' is widely used in cosmology, in addition to the usual 
``metric formalism'' \cite{Palatini}. Furthermore, if the matter 
part of the action $S^{(m)}$ also depends on the connection, one 
obtains a third possibility, metric-affine gravity theories 
\cite{Sotiriou,LiberatiSotiriou}. In the following we consider 
the metric approach to modified  gravity but the methods and the 
conclusions apply to the Palatini approach as well. 

We briefly recall the dynamical equivalence 
between $f(R)$ gravity and scalar-tensor gravity 
\cite{TeyssandierTourrenc,Wands,ChibaPLB03} (for the dynamical equivalence between scalar-tensor theories see 
Ref.~\cite{TorresVucetich}). By introducing 
an auxiliary field $\phi$ the action (\ref{1}) becomes
\begin{equation} \label{2}
S=\frac{1}{2\kappa}\int d^4x \, \sqrt{-g} \left[ f( 
\phi) +\frac{df}{d\phi} \left( R-\phi \right) \right] +S^{(m)}
\end{equation}
if $d^2f/d\phi^2\neq 0$. This action integral can be written as 
\begin{equation} \label{3}
S=\frac{1}{2\kappa}\int d^4x \, \sqrt{-g} \left[ 
\psi (\phi) R -V(\phi) \right] +S^{(m)} \;,
\end{equation}
where
\begin{equation} \label{3bis}
\psi(\phi) \equiv \frac{df}{d\phi} \;, \;\;\;\;\;\;
V(\phi)\equiv \phi \, \frac{df}{d\phi} -f(\phi) \;.
\end{equation}
This action describes  a scalar-tensor theory of gravity 
(\cite{BransDicke,ST} --- see \cite{mybook} for 
a review)   with 
Brans-Dicke parameter $\omega=0$. The corresponding field 
equations are
\begin{eqnarray} 
G_{ab} &=& \frac{1}{\psi} \left( \nabla_a \nabla_b \psi-g_{ab} 
\, \Box\psi -\frac{V}{2}\, g_{ab} \right) +\frac{\kappa}{\psi} \, 
T_{ab}^{(m)} \;, 
\label{ST1} 
\\
&& \nonumber \\
& \; \; & R\, \frac{d\psi}{d \phi}-\frac{dV}{d\phi}=0 \;. 
\label{ST2}
\end{eqnarray}
Trivially, if $\phi=R$, the action (\ref{2}) reduces to 
(\ref{1}). Vice-versa, by varying the action (\ref{2}) with 
respect to $\phi$ and assuming that $S^{(m)}$ is independent of 
$\phi$, one obtains 
\begin{equation} \label{phi=R}
\left( R-\phi \right) f''(\phi)=0 \;,
\end{equation}
which yields $\phi=R$ provided that $f'' \neq 0$ (a prime now 
denotes 
differentiation with respect to $\phi$). Similarly, one 
shows that $f(R)$ gravity in the Palatini formalism is 
equivalent to a $\omega=-3/2$ Brans-Dicke theory when the 
matter action is independent of the connection 
(e.g., \cite{Sotiriou}).
This dynamical 
equivalence between theories can be quite useful but it should 
not be abused. It has been used to constrain $f(R)$ gravity 
based on Solar System bounds on the post-Newtonian parameters of 
scalar-tensor gravity \cite{ChibaPLB03,PPN,SotiriouPPN}. The 
underlying logic is that deviations from general relativity (GR) 
are not detected in our local 
spacetime neighborhood, therefore these deviations (if they 
exist) must be small.  This ``closeness of $f(R)$ gravity to 
GR'' implies that $f(R)$ gravity reduces to GR  
in an appropriate limit, which we address here. While 
there is in principle no problem 
in taking this limit directly in modified gravity, the dynamical 
equivalence with 
scalar-tensor gravity breaks down in this limit. In fact, GR  
corresponds to $f(R)=R$ and $f'' \equiv 0$, 
while the dynamical equivalence {\em requires} $f''\neq 0$. This 
fact has been overlooked  and the dynamical equivalence has been 
used beyond its realm of validity in the limit to GR by 
advocating  the parametrized post-Newtonian (PPN) formalism of 
scalar-tensor gravity.
This procedure is invalid and it explains why opposite claims 
of compatibility/non-compatibility of $f(R)$ gravity with Solar 
System experiments occur in the literature - worse, even the 
Newtonian limit is the subject of dispute \cite{PPN,SotiriouPPN}. 

The limit to GR of the equivalent scalar-tensor 
theory is more general than the weak field limit: it includes 
the strong field regime and 
it turns out to be a singular limit, as shown below. Moreover, 
the limit of 
Brans-Dicke theory to GR in vacuum, or in the 
presence of ``conformal'' matter (i.e., with vanishing trace of 
the stress-energy tensor), is riddled with 
problems. Therefore, one 
must be particularly careful in basing all of one's 
conclusions on the compatibility with Solar System experiments 
(which test the gravitational field in vacuo or at very low 
densities) solely on the equivalence 
with scalar-tensor gravity. In the following section we discuss 
the 
direct limit of $f(R)$ gravity to GR without 
using the equivalence with scalar-tensor gravity, and then the 
corresponding limit to GR of the equivalent scalar-tensor 
gravity, showing the problems arising in this last situation. We 
do not want to commit ourselves to specific choices of the 
function $f(R)$ (e.g., the CDTT model \cite{MG1,CDTT}, etc.) but 
we consider a general form of the function $f(R)$.

\section{The limit of $f(R)$ gravity to general relativity}

Perhaps the easiest way to consider the limit of $f(R)$ gravity 
to Einstein's theory consists of introducing a small parameter 
$\epsilon$ such that $f(R)$ can be expressed as 
\cite{footnote1} 
\begin{equation} \label{4}
f(R)=R+\epsilon \, \varphi(R) \;.
\end{equation}
The action of GR $ S_{GR}= \left( 2\kappa \right)^{-1}\int d^4 x 
\sqrt{-g} \, R +S^{(m)} $ is obtained in the limit 
$\epsilon\rightarrow 
0$ \cite{footnote2}. The field equations become 
\begin{eqnarray} 
 \left( 1+\epsilon \varphi ' \right) R_{ab} -\frac{1}{2}\left( 
R+\epsilon \varphi \right) g_{ab} &= & \epsilon \nabla_a\nabla_b 
\varphi' -\epsilon \, g_{ab}\Box \varphi' \nonumber \\
&&\nonumber \\
&& +\kappa \, T_{ab}^{(m)} \label{5} 
\end{eqnarray}
which, in the limit $\epsilon \rightarrow 0$, formally reduce to 
the Einstein equations $G_{ab} =\kappa T_{ab}^{(m)}$. So, there 
is no problem in taking the limit of $f(R)$ gravity to GR 
directly. Let us consider now the ``equivalent'' scalar-tensor 
theory (\ref{3}). Although the conventional way to obtain this 
limit is letting the Brans-Dicke parameter $\omega$ go to 
infinity, 
here $\omega $ is fixed to be zero. The limit to GR can again be 
obtained by letting $\epsilon $ go to zero in the equations of 
scalar-tensor theory, which is equivalent to taking the limit 
$\phi \rightarrow $~constant. Assuming that $f(R)$ is given by 
eq.~(\ref{4}), the non-linear theory (\ref{1}) is equivalent to 
(\ref{2}) with
\begin{equation} \label{6}
\psi(\phi) =1+\epsilon \, \varphi'(\phi) \;, 
\;\;\;\;\;\;\;\;\;\;
V(\phi)=\epsilon \left( \varphi'\phi-\varphi \right)
\end{equation}
provided that $f''\neq 0$. Now, in the limit $\epsilon 
\rightarrow 0$, ~~$f''=\epsilon \, \varphi'' \rightarrow 0$ and 
the 
equivalence is broken. Formally, the field equation (\ref{ST1}) 
reduces to the Einstein equation while (\ref{ST2}) is identically 
satisfied. There are however, problems with this procedure. One 
should also find the asymptotic behaviour of the field $\phi$ as 
$\epsilon \rightarrow 0$. The situation is analogous to the 
standard limit to GR of Brans-Dicke theory, the prototype of 
scalar-tensor theories, in which GR is 
usually obtained by taking the limit $\omega\rightarrow \infty$. 
The standard textbook 
presentation provides the asymptotic behaviour of the 
Brans-Dicke field $\phi_{BD}$:
\begin{equation} \label{7}
\phi_{BD}=\phi_0+\mbox{O}\left( \frac{1}{\omega} \right) \;,
\end{equation}
where $\phi_0 $ is a constant \cite{Weinberg}. But when the 
trace of the energy-momentum tensor of matter $T^{(m)}$ 
vanishes, the $\omega\rightarrow \infty$ limit fails to give 
back GR --- this phenomenon  is reported for a number of exact 
Brans-Dicke solutions \cite{exactBDsolutions} and is  
identified as a general feature of Brans-Dicke theory explained 
by a 
restricted conformal invariance enjoyed by the theory when 
$T^{(m)}=0$ \cite{VFBDlimit}. This anomalous behaviour is 
intimately linked with the asymptotics displayed by the 
Brans-Dicke field in these situations  
\cite{footnote3,BanerjeeSen,VFBDlimit} 
\begin{equation} \label{8}
\phi_{BD}=\phi_0+\mbox{O}\left( \frac{1}{\sqrt{\omega}} \right) 
\;\;\;\;\;\;\;\;\;\;\;\;\;\; \left( T^{(m)}=0 \right) .
\end{equation}
Similarly, the examination of the asymptotics of the scalar 
field $\phi$ as the parameter $\epsilon$ tends to zero should 
provide a useful check of the limiting procedure.  In the 
scalar-tensor  equivalent of $f(R)$ gravity, the 
Brans-Dicke parameter is fixed  to be $\omega=0$ in the metric 
formalism ($\omega=-3/2$ in 
the Palatini formalism) and we must necessarily come up with a  
different way of taking the limit to GR, hence the possibility 
considered of letting $\epsilon $ going to zero while 
$\phi$ becomes constant. In this case, we 
should obtain a reasonable asymptotic behaviour for the fields 
$g_{ab}$ and $\phi$, say
\begin{equation} \label{9}
g_{ab}=g_{ab}^{(GR)}+\epsilon \, h_{ab} \;, 
\;\;\;\;\;\;\;\;\;\;
\phi=\phi_0+r(\epsilon) \;,
\end{equation}
where $g_{ab}^{(GR)} $ is the general-relativistic 
metric, $\phi_0$ is a constant, and the remainder $r(\epsilon)$ 
tends to zero as $\epsilon \rightarrow 0$. However, this is not 
the case. By inserting eq.~(\ref{4}) into eq.~(\ref{ST1})  one obtains
\begin{eqnarray} 
 G_{ab} &=& \frac{\epsilon}{1+\epsilon\, \varphi'} \left[ 
\nabla_a\nabla_b  \varphi' -g_{ab} \, \Box \varphi' 
+\frac{1}{2}  \left( \varphi  -\phi_0\varphi' \right)g_{ab} 
\right] \nonumber \\
&&\nonumber \\
&& +\frac{ \kappa \, T_{ab}^{(m)}}{1+\epsilon \, \varphi'} 
\;,\label{10}
\end{eqnarray}
while
\begin{eqnarray}
 \Box \phi &= & -\frac{\epsilon \varphi'''}{1+\epsilon \varphi'} 
\, \nabla^c\phi\nabla_c\phi +\frac{ \epsilon \left( 
\phi\varphi'-2\varphi\right)-\phi}{3\epsilon 
\varphi''} \nonumber \\
&&\nonumber \\
&& +\frac{1+\epsilon \, \varphi'}{\epsilon \, \varphi''}\,  
\kappa \, T^{(m)} 
\;,\label{11}
\end{eqnarray}
where the indices are raised and lowered with $g_{ab}^{(GR)}$. 
Further substitution 
of eq.~(\ref{9}) yields, in the limit 
$\epsilon\rightarrow 0$,
\begin{equation} \label{12}
r(\epsilon) =\mbox{O}\left( \frac{1}{\epsilon} \right) \;.
\end{equation}
The remainder $r(\epsilon) $ diverges instead of vanishing: 
$\epsilon \rightarrow 0$  is  a singular limit of the 
``equivalent'' scalar-tensor version of $f(R)$ gravity while 
the direct limit $\epsilon \rightarrow 0$ of $f(R)$ gravity 
does not lead to this problem. Again, this reflects the 
breakdown of the dynamical equivalence in the limit to GR in 
which $f'' \rightarrow 0$.  Note that the procedure 
employed here 
parallels the procedure used to obtain the estimate (\ref{7}) 
for Brans-Dicke theory \cite{Weinberg, BanerjeeSen, Chauvineau}.

\section{Stability of de Sitter space in modified and 
scalar-tensor gravity}

In this section we consider the cosmological dynamics of 
modified gravity. By assuming the spatially flat 
Friedmann-Lemaitre-Robertson-Walker (FLRW) metric
\begin{equation} \label{FLRW}
ds^2=-dt^2+a^2(t)\left( dx^2+dy^2+dz^2 \right) 
\end{equation}
in comoving coordinates $\left( t,x,y,z \right) $, the field 
equations (\ref{1bis}) of modified gravity  reduce to
\begin{eqnarray}
H^2 &=& \frac{ 1}{3 f'(R)} \left\{  
\frac{f(R)- R f'(R)}{2} -3H\dot{R}f''(R) \right. 
\nonumber\\
&&\nonumber \\
&& \left. + \kappa \, \rho^{(m)}\right\}  \;, \\
&&\nonumber \\
2\dot{H}+3H^2 &= & -\frac{1}{f'(R)} \left\{ \left( 
\dot{R}\right)^2 f'''(R) 
+2H\dot{R} f''(R) \right.\nonumber \\
&& \nonumber \\
&&\left. +\ddot{R} f''(R)  -\frac{1}{2} \left[ 
f(R)-R f'(R) \right] + \kappa \, P^{(m)} \right\} \;.\nonumber\\
&& 
\end{eqnarray}
Consider, for simplicity, the situation in which the curvature 
terms dominate the dynamics and $\rho^{(m)}$ and $P^{(m)}$ are 
negligible.  Then, the scale factor $a(t)$ enters the field 
equations only through the Hubble parameter $H\equiv \dot{a}/a$ 
and it is natural to use $H$ as the dynamical variable. The 
field equations are of fourth order in $a$ (hence the name 
``fourth order gravity'' given to $f(R)$ theories) and of third 
order in $H$. The main result of this section is that the 
$\omega=0$  scalar-tensor theory does not provide the same 
stability  condition derived in $f(R)$ gravity, but two 
different ones according to the type of perturbations 
considered. Therefore,  these two theories are inequivalent 
with regard to stability.

The equilibrium points in the phase space $\left( H, \dot{H}, 
\ddot{H} \right) $ are de Sitter spaces characterized by 
constant Hubble parameter $H_0$ given by
\begin{equation}
H_0^2=\frac{f_0}{6f_0'}
\end{equation}
and $R_0=12H_0^2$. The stability of these de Sitter 
spaces with 
respect to both homogeneous and inhomogeneous perturbations was 
studied in Ref.~\cite{VFrapid}, with the result that the 
stability conditions with respect to both types of 
perturbations coincide and, in our notations, are expressed by 
\cite{footnote4}
\begin{equation} \label{SC1}
\frac{ \left( f_0' \right)^2 -2f_0 \, f_0''}{f_0' \, f_0''}\geq 0 
\;.
\end{equation}
The study of stability with respect to inhomogeneous 
perturbations, which 
are inherently gauge-dependent, requires the use of a 
gauge-invariant formalism. It is counterintuitive that the  
stability condition with respect to inhomogeneous perturbations 
is not more restrictive than the corresponding stability 
condition with respect to homogeneous perturbations. This is not 
the case, for example, in scalar-tensor gravity. Stability in  
scalar-tensor gravity was also studied in 
Refs.~\cite{VFrapid,VFMJ}, but the cases $\omega=0 $ and 
$\omega=-3/2$ were excluded by the analysis. In the following 
subsections we  study the stability of the de Sitter equilibrium 
points in the $\omega= 0$ equivalent of metric  $f(R)$ gravity. 
As expected, the stability condition with respect to 
inhomogeneous perturbations turns out to be different from the 
stability condition with respect to 
homogeneous ones.

\subsection{Stability with  respect to homogeneous 
perturbations in  $\omega=0$ scalar-tensor gravity}

In the $\omega=0$ scalar-tensor theory described by the action 
(\ref{2}) the field equations become, in the metric 
(\ref{FLRW}) and in the absence of matter,
\begin{eqnarray}
3 f' H^2 &=&    \frac{1}{2} \left( \phi f' 
-f \right) -3 H f'' \dot{\phi}  \;,  \label{SS24} \\
&&\nonumber \\
-2 f' \dot{H} &=& f''' \left( \dot{\phi}\right)^2 +  f'' 
\ddot{\phi}   - H f'' \dot{\phi}  \;, \label{SS25}\\
&&\nonumber \\
&& f'' R-V'=0 \;.
\end{eqnarray}
Manipulation of eqs.~(\ref{SS24}) and (\ref{SS25}) leads to the 
Klein-Gordon-like equation for $\phi$
\begin{equation} 
\ddot{\phi} +3H\dot{\phi}=\frac{1}{3f''}\left[ -f' R -3f''' 
\left( \dot{\phi} \right)^2 +2\left( \phi f' - f \right) \right] 
\; .\label{S27}
\end{equation}
The equilibrium points in this picture correspond to de 
Sitter spaces with constant scalar field given by
\begin{equation}
H_0^2=\frac{f_0}{6f_0'} \;, \;\;\;\;\;\;\;\;\;\;
\phi_0=\frac{2f_0}{f_0'}=R_0 \;.
\end{equation}
Homogeneous perturbations of the de Sitter fixed points are 
described by
\begin{equation}
H(t)=H_0+\delta H(t) \;, \;\;\;\;\;\;\;\;\;\;
\phi(t)=\phi_0+\delta \phi(t) \;,
\end{equation}
and obey the evolution equations
\begin{eqnarray}
&& 12H_0 f_0' \, \delta H +\left( 6H_0^2-\phi_0 
\right)f_0'' \, \delta  \phi +6H_0 f_0'' \, \delta \dot{\phi} =0 
\;, \\
&&\nonumber \\
&& -2 f_0'\, \delta \dot{H} = f_0'' \, \delta \ddot{\phi} -H_0 
f_0'' \, \delta \dot{\phi} \\
&& \nonumber \\
&& 6f_0''\delta \dot{H}+24 H_0f_0''\delta H+\left( 12H_0^2 
f_0'''-\phi_0 f_0''' -f_0'' \right)\delta\phi =0 \;,\nonumber \\
&& 
\end{eqnarray}
where the constraint $\phi=R$ has been used. By eliminating 
$\delta \dot{H} $ and using the zero-order equations for de 
Sitter 
space, one obtains
\begin{equation}
\delta \ddot{\phi}+3H_0 \, \delta \dot{\phi} +\frac{1}{3f_0' 
\left( 
f_0'' \right)^2} \left[ \left( f_0' \right)^2-2f_0 \right] \delta 
\phi =0 \;.
\end{equation}
Stability is achieved, and the perturbations $\delta \phi$ do 
not run away, when the effective mass squared in this harmonic 
oscillator equation is non-negative, i.e., 
\begin{equation} \label{SC2}
\frac{ \left( f_0' \right)^2 -2f_0}{ f_0' } \geq 0 
\;.
\end{equation}
This is the desired stability condition with respect to 
homogeneous perturbations in the $\omega=0$ scalar-tensor theory 
that is supposed to be equivalent to $f(R)$ gravity. This 
condition is different from (\ref{SC1}) showing that, at best, 
the equivalence should be treated with care. The stability 
condition (\ref{SC2}) should be compared with the stability 
condition with respect to inhomogeneous perturbations, which we 
derive in the next subsection.

\subsection{Inhomogeneous perturbations in $\omega=0$ 
scalar-tensor gravity}

The analysis of inhomogeneous perturbations necessarily requires  
a gauge-independent formalism. We adopt the covariant and 
gauge-invariant formalism of Bardeen-Ellis-Bruni-Hwang 
\cite{Bardeen} in the  version of Hwang  valid for 
generalized gravity \cite{Hwang}. The metric perturbations are 
given by
\begin{eqnarray} 
g_{00} & = & -a^2 \left( 1+2AY \right) \;, \\
&& \nonumber \\
g_{0i} & = & -a^2 \, B \, Y_i  \;,  \\
&& \nonumber \\
g_{ij} & =& a^2 \left[ h_{ij}\left(   1+2H_L \right) +2H_T \, 
Y_{ij}  \right] \;,
\end{eqnarray}
where $Y$, $Y_{i} $, and $ Y_{ij}$ are the scalar, 
vector, and tensor harmonics, respectively, 
satisfying
\begin{equation} 
\bar{\nabla_i} \bar{\nabla^i} \, Y =-k^2 \, Y \;, \;\;\;\;\;\;\;
Y_i= -\frac{1}{k} \, \bar{\nabla_i} Y \;,
\end{equation}
\begin{equation} 
 Y_{ij}= \frac{1}{k^2} \, \bar{\nabla_i}\bar{\nabla_j} Y 
+\frac{1}{3} \, Y \, h_{ij} \;.
\end{equation}
Here $h_{ij} $ is the three-dimensional metric of the FLRW 
background and the operator $ \bar{\nabla_i} 
$ is the covariant derivative associated with  $h_{ij}$, while 
$k$ is an eigenvalue. The gauge-invariant variables 
used are Bardeen's potentials and the 
 Ellis-Bruni  variable 
\begin{equation} 
\Phi_H = H_L +\frac{H_T}{3} +\frac{ \dot{a} }{k} \left( 
B-\frac{a}{k} \, \dot{H}_T \right) \;, 
\end{equation}

\begin{equation} 
 \Phi_A = A  +\frac{ \dot{a} }{k} \left( B-\frac{a}{k} \, 
\dot{H}_T \right)
+\frac{a}{k} \left[ \dot{B} -\frac{1}{k} \left( a \dot{H}_T \right)\dot{}  \right] \;, 
\end{equation}

\begin{equation}
 \Delta \phi = \delta \phi  +\frac{a}{k} \, \dot{\phi}  \left( 
B-\frac{a}{k} \, \dot{H}_T 
\right) 
\; ,
\end{equation}
with  $ \Delta f$ and $\Delta R$ defined similarly to the last 
equation. The first order equations of motion for the 
gauge-invariant perturbations can be found in Ref.~\cite{Hwang}; 
they simplify considerably in the de Sitter background, reducing 
to

\begin{equation}  \label{S22}
\Delta \ddot{\psi} + 3H_0  \Delta \dot{\psi} 
+ \left( \frac{k^2}{a^2}- 4H_0^2 \right) \Delta \psi 
+\frac{\psi_0}{3}\, \Delta R =0 \;,
\end{equation}

\begin{equation} \label{S23}
-\dot{\Phi}_H + H_0 \Phi_A =\frac{1}{2} \left( \frac{\Delta 
\dot{\psi}}{\psi_0} -H_0 \, \frac{ \Delta \psi}{\psi_0} \right) \;,
\end{equation}

\begin{equation}  \label{S24}
\left( \frac{k}{a} \right)^2 \Phi_H= -\frac{k^2}{2a^2} \,  
\frac{\Delta \psi}{\psi_0}  \;,
\end{equation}

\begin{equation}   \label{S25}
\Phi_A + \Phi_H =  - \frac{\Delta \psi}{\psi_0} \; ,
\end{equation}

\begin{equation} \label{S26}
\ddot{H}_T +3H_0  \, \dot{H}_T +\frac{k^2}{a^2} \, H_T=0 \;,
\end{equation}

\begin{eqnarray} 
&& \ddot{\Phi}_H + 3H_0 \dot{\Phi}_H  - H_0  \dot{\Phi}_A 
-\frac{V_0}{2\psi_0}\, \Phi_A  \nonumber \\
&&\nonumber \\
&& =  - \frac{1}{2} \left[ \frac{  \Delta \ddot{\psi}}{\psi_0} 
+2H_0 \,  \frac{\Delta \dot{\psi}}{\psi_0} 
- \frac{V_0}{2\psi_0^2 } \,  \Delta \psi \right]  \; ,
\end{eqnarray}

and
\begin{eqnarray} 
\Delta R &=& 6 \left[ \ddot{\Phi}_H + 4H_0 \dot{\Phi}_H + 
\frac{2}{3} \frac{k^2}{a^2} \, \Phi_H  -H_0 \dot{\Phi}_A 
\right.\nonumber\\
&&\nonumber \\
&& \left. + \left(  \frac{k^2}{3a^2} -4H_0^2 \right) \Phi_A 
\right] \;, \label{S28}
\end{eqnarray}
where $a=a_*\, \mbox{e}^{H_0 t} $ is the unperturbed scale 
factor. Equation~(\ref{S26}) exhibits a positive effective mass 
squared $ k^2/a^2$ for the tensor modes $H_T$, therefore de 
Sitter spaces are always stable with respect to these modes, to 
linear order. By using eq.~(\ref{S25}) to eliminate $\Phi_A$ and 
Taylor-expanding $\Delta \psi=\psi_0'\, \Delta \phi+\, ...\;$, 
one  easily obtains
\begin{equation}  
\Phi_H  = \Phi_A= -  \frac{ \psi_0'}{2\psi_0}\, \Delta \phi  
\end{equation}
and
\begin{equation}
\Delta R =-\frac{3\psi_0'}{\psi_0} \left[ \Delta 
\ddot{\phi}+3H_0\Delta \dot{\phi}+\left( \frac{k^2}{a^2}-4H_0^2 
\right) \Delta \phi \right] \;.
\end{equation}
By using the fact that $\Delta R=\Delta \phi$, the equation for 
the scalar perturbations $\Delta\phi$ is obtained:
\begin{equation}
\Delta \ddot{\phi}+3H_0\Delta \dot{\phi} +\left( \frac{k^2}{a^2} 
-4H_0^2 +\frac{\psi_0}{3\psi_0'} \right) \Delta \phi=0 \;.
\end{equation}
The term $\left( k/a \right)^2$ can be neglected at late times 
when $a$ diverges exponentially fast and the stability condition 
for the perturbations $\Delta \phi$ (and $\Phi_H=\Phi_A\propto 
\Delta \phi$) is therefore
\begin{equation}\label{SC3}
\frac{ \left(f_0'\right)^2-2f_0 \, f_0''}{f_0'  \, f_0''} \geq0 
\;.
\end{equation}
This is the desired  stability condition with respect to 
inhomogeneous perturbations. It is different from the 
condition  for homogeneous perturbations (\ref{SC2}) and it 
coincides with  (\ref{SC1}). Not only one of the 
stability conditions with  respect to homogeneous and 
inhomogeneous perturbations  (\ref{SC2}) and (\ref{SC3}) of the 
$\omega=0$ scalar-tensor 
``equivalent'' of $f(R)$ gravity fails to coincide with the 
condition (\ref{SC1}), but they also differ from each other. The 
purported equivalence is not a true equivalence. The reason 
should be looked for in the fact that, while in a true 
scalar-tensor theory the Brans-Dicke-like field $\phi$ is a true 
dynamical field whose evolution is only ruled by the dynamical 
field equations, in the theory considered here $\phi$ is forced 
to obey the additional constraint $\phi=R$, thus limiting its 
natural evolution. In other words, we have gone from fourth order 
equations to second order equations by adding  a scalar degree of 
freedom to the theory, which has now spin~0 content in addition 
to spin~2, but the scalar degree of freedom is somehow 
constrained by the condition $\Delta \phi=\Delta R$.

\section{Discussion and conclusions}

While the limit to GR $\epsilon \rightarrow 0$ in the action 
(\ref{1}) of modified gravity presents no problems of principle, 
the limit to GR of the ``equivalent'' scalar-tensor theory is 
ill-defined. In view of the problems presented by this limit in 
Brans-Dicke theory, special care is advised when using the 
dynamically equivalent scalar-tensor theory to analyse the weak 
field limit of modified gravity. As shown above, the limit to 
GR  of the ``equivalent'' $\omega=0$ Brans-Dicke theory is a 
singular one, illustrating the fact that the equivalence breaks 
down in this limit. {\em A posteriori} it is easy to see that 
this is implied by the fact that $f''(R) \rightarrow 0$ in this 
limit. 

Another issue is the following: if the dynamical equivalence 
were to hold in the limit to GR, the experimental bound $\omega > 
40000 $ 
\cite{Bertotti} would be in violent conflict with the values of 
the Brans-Dicke parameter $\omega=0$ or $-3/2$ obtained, unless 
the field 
$\phi$ is short-ranged. The effective mass of $\phi$ is given by
\begin{equation}
m_{eff}^2 =V''=\epsilon \left( \varphi''+\phi \, \varphi ''' 
\right)
\end{equation}
and vanishes as $\epsilon \rightarrow 0$, making it impossible 
to suppress the violation of the bounds on $\omega$, for {\em 
any} form of the function $f(R)$. This contradicts the results of 
Refs.~\cite{MGviable,SotiriouPPN} which support the viability of 
the weak 
field limit of the theory for 
specific forms of $f(R)$. As it turns out, the PPN limit of the 
associated scalar-tensor theory bears no relation to the weak 
field limit of the original modified gravity theory.

As a consequence, the conclusion \cite{ChibaPLB03,PPN} that 
modified gravity always violates the stringent Solar System 
bounds on scalar-tensor gravity \cite{Bertotti} is 
invalid. The issue of the correct Newtonian and post-Newtonian 
limit of such theories is still open and should be approached 
directly without using the dynamical equivalence discovered in 
Refs.~\cite{TeyssandierTourrenc,Wands}, which is still useful 
for 
other purposes. The regime that is more interesting, however, is  
the one in which the deviations from GR in the Solar 
System are small. A complete study of the PPN limit of 
general modified gravity scenarios without resorting to the 
equivalent scalar-tensor theory (initiated in 
Refs.~\cite{SotiriouPPN,ShaoCaiWangSuPLB}) will be presented 
elsewhere.

Another apparent problem with the limit to GR lies in the fact 
that 
$\phi$ must become constant: because $\phi=R$, were this 
limit correct, it could only  reproduce solutions with constant 
Ricci curvature (which includes vacuum solutions and solutions 
sourced by conformal matter). Although this could work for 
vacuum solutions used to describe Solar System experiments, it 
is by no means acceptable to have a limit to GR valid only 
for special solutions: the limit must apply to the general 
theory.  However, we believe that this second problem is not 
very 
relevant  because, when  $f''$ vanishes in eq.~(\ref{phi=R}) 
in this limit,   the equality $\phi=R$ is no longer enforced.

Finally, the viability of modified gravity scenarios does not 
rely only on its  correct weak field limit: other issues are the 
presence of ghosts and instabilities  and, of 
course, a correct  cosmological dynamics. These have been 
considered separately in  the literature \cite{instabilities}.

\begin{acknowledgments}
This work was supported by the Natural Sciences and Engineering 
Research Council of Canada ({\em NSERC}).
\end{acknowledgments}


\begin{thebibliography}{99}

\bibitem{SN} A.G. Riess {\em et al.}, {\em Astron. J.} 
{\bf 116}, 1009 (1998); {\em Astron. J.} {\bf 118}, 2668 
(1999);
{\em Astrophys. J.} {\bf 560}, 49 (2001);
{\em Astrophys. J.} {\bf 607}, 665 (2004);
S. Perlmutter {\em et al.}, {\em Nature} {\bf 391}, 51 (1998);
{\em Astrophys. J.} {\bf 517}, 565 (1999); 
J.L. Tonry {\em et al.}, {\em Astrophys. J.} {\bf 594}, 
1 (2003);
R.  Knop {\em et al.}, {\em Astrophys. J.} {\bf 598}, 102 
(2003);
B. Barris {\em et al.}, {\em Astrophys. J.} {\bf 602}, 571 (2004).

\bibitem{CMB} 
A.D. Miller {\em et al.}, {\em Astrophys. J. Lett.} {\bf 524}, 
 L1 (1999); 
P. de Bernardis {\em et al.}, {\em Nature} {\bf 404}, 955 
(2000);
A.E. Lange {\em et al.}, {\em Phys. Rev. D} {\bf 63}, 042001 (2001);
A. Melchiorri, L. Mersini, C.J. Odman and M. Trodden, {\em 
Astrophys. J. Lett.} {\bf 536},   
L63 (2000); 
S. Hanany {\em et al.}, {\em Astrophys. J. Lett.} {\bf 545},  
L5 (2000); 
D.N. Spergel {\em et al.}, {\em Astrophys. J. (Suppl.)} {\bf 
148},  175 (2003);
C.L. Bennett {\em et al.}, {\em Astrophys. J. (Suppl.)} {\bf 
148}, 1 (2003);
T.J. Pearson {\em et al.}, {\em Astrophys. J. } 
{\bf 591}, 556 (2003);
A. Benoit {\em et al.},  {\em Astron. Astrophys.} 
{\bf 399}, L25  (2003). 

\bibitem{w} J.S. Alcaniz, {\em Phys. Rev. D} {\bf 69}, 083521 
(2004);
A. Melchiorri, L. Mersini, C.J. Odman, and M. Trodden, {\em Phys. 
Rev. D} {\bf 68}, 043509 
(2003);
S. Hannestad and E. Mortsell, {\em Phys. Rev. D} {\bf 66}, 
023510 (2002); 
U. Alam {\em et al.}, {\em Mon. Not. R. Astr. Soc.} {\bf 354}, 
275 (2004);
S. Nesseris and L. Perivolaropoulos, {\em Phys. Rev. D} {\bf 
70}, 043531 (2004); 
T.R. Choudhury and T. Padmanabhan, {\em Astron. Astrophys.} 
{\bf 429},  807 (2005).

\bibitem{phantom} 
S. Capozziello, S. Nojiri and S.D. Odintsov, {\em Phys. Lett. B} 
{\bf 632}, 597 (2006);
S. Nojiri and S.D. Odintsov, hep-th/0506212; 
{\em Phys. Rev. D} {\bf 72}, 023003 (2005);
V. Faraoni, {\em Class. Quant. Grav.} {\bf 22}, 
3235 (2005); 
W. Fang {\em et al.}, {\em Int. J. Mod. Phys. D} {\bf 15}, 199 
(2006); 
P.F. Gonzalez-Diaz and J.A. Jimenez-Madrid, {\em Phys. 
Lett. B} {\bf 596},  16 (2004);
M.G. Brown, K. Freese and W.H. Kinney, astro-ph/0405353;
E. Elizalde, S. Nojiri and S.D. Odintsov
 {\em Phys. Rev. D} {\bf 70},  043539 (2004); 
{\em Phys. Lett. B} {\bf 574}, 1 (2003); 
{\em Phys. Rev. D} {\bf 70}, 043539 (2004);  
J.-G. Hao  and  X.-Z.  Li, {\em Phys. Lett. B} {\bf 606},  
 7 (2005);
{\em Phys. Rev. D} {\bf 68}, 043501 (2003);
{\em Phys. Rev. D} {\bf 69}, 107303 (2004);
J.M. Aguirregabiria, L.P. Chimento and R. Lazkoz, {\em 
Phys. Rev. D} {\bf 70},  023509 (2004);
Y.-S.  Piao  and  Y.-Z. Zhang, {\em Phys. Rev. D} {\bf 70}, 
063513 (2004);
P.F. Gonzalez-Diaz, {\em Phys. Rev. D} {\bf 68}, 021303(R) 
(2003);
{\em Phys. Rev. D} {\bf 69},  063522 (2004);
{\em Phys. Lett. B} {\bf 586},  1 (2004);
H.Q. Lu, {\em Int. J. Mod. Phys. D} {\bf 14}, 355 (2005);
V.B. Johri, {\em Phys. Rev. D} {\bf 70}, 041303 (2004);
H. Stefanci\'{c}, {\em Phys. Lett. B} {\bf 586},  5 (2004);
D.J. Liu and  X.Z. Li, {\em Phys. Rev. D} {\bf 68},  
067301 (2003);
J.G. Hao and X.Z.  Li, {\em Phys. Rev. D} {\bf 69},  
107303 (2004);
M.P. Dabrowski, T. Stachowiak and M. Szydlowski, 
  {\em Phys. Rev. D} {\bf 68},  103519 (2003);
E. Babichev, V. Dokuchaev and  Yu. Eroshenko,  {\em Phys. 
Rev. Lett.}  {\bf 93}, 021102 (2004);
Z.K. Guo, Y.S. Piao and Y.Z. Zhang, 
  {\em Phys. Lett. B} {\bf 594}, 247 (2004);
J.M. Cline, S.  Jeon and G.D. Moore,  
  {\em Phys. Rev. D} {\bf 70}, 043543  (2004); 
S. Nojiri  and S.D. Odintsov, {\em Phys. Lett. B} {\bf 562},  
147 (2003);
V. Faraoni, {\em Ann. Phys. (NY)} {\bf 317}, 366 (2005); 
V. Faraoni, M.N. Jensen, and S.A. Theuerkauf, gr-qc/0605050;
L. Mersini, M. Bastero-Gil and P. Kanti, 
  {\em Phys. Rev. D} 64 (2001) 043508; 
M. Bastero-Gil, P.H. Frampton  and L. Mersini, 
  {\em Phys. Rev. D} 65 (2002) 106002; 
P.H. Frampton, {\em Phys. Lett. B} 555 (2003) 139.

\bibitem{MG1} S. Capozziello, S. Carloni and A. Troisi, 
astro-ph/0303041.

\bibitem{CDTT} S.M. Carroll, V. Duvvuri, M. Trodden and M.S. 
Turner, {\em Phys. Rev. D} {\bf 70}, 043528 (2004).

\bibitem{Mtheory} S. Nojiri and S.D. Odintsov, {\em Phys. Lett. 
B} {\bf 576}, 5 (2003); S. Nojiri, S.D. Odintsov, and M. Sami, 
hep-th/0605039.

\bibitem{MG3} 
S. Capozziello, V.I. Cardone, S. Carloni and A. Troisi, {\em 
Int. J. Mod. Phys. D} {\bf 12}, 1969 (2003);
S. Nojiri and S.D. Odintsov, {\em Phys. Rev. D} {\bf 68}, 123512 
(2003);
D.N. Vollick, {\em Phys. Rev. D} {\bf 68}, 063510 (2003)
S. Carloni, P.K.S. Dunsby, S. Capozziello and A. Troisi, {\em 
Class. Quant. Grav.} {\bf 22}, 4839 (2005);
D.A. Easson, {\em Int. J. Mod. Phys. A} {\bf 19}, 5343 (2004);
D.A. Easson, F.P. Schuller, M. Trodden and 
M.N.R. Wohlfarth, {\em Phys. Rev. D} {\bf 72}, 
043504 (2005);
G.J. Olmo and W. Komp, gr-qc/0403092;
\`{E}.\`{E}. Flanagan, {\em Phys. Rev. Lett.} {\bf 92}, 071101 
(2004);
{\em Class. Quant. Grav.} {\bf 21}, 417 
(2004); {\bf 21}, 3817 (2004); 
M. Ishak, A. Upadhye and D.N. Spergel, astro-ph/0507184; S. 
Nojiri and S.D. Odintsov, {\em Phys. Lett. B} {\bf 576}, 5 
(2003); {\bf 599}, 137 (2004);
G. Allemandi, A. Borowiec and M. Francaviglia, {\em Phys. Rev. D} 
{\bf 70}, 103503 (2004);
A. Lue, R. Scoccimarro and G. Starkman, {\em Phys. Rev. D} {\bf 
69}, 044005 (2004); 
T. Koivisto, gr-qc/0505128;
M. Sami, A. Toporensky, P.V. Tretjakov and S. Tsujikawa, {\em 
Phys. Lett. B} {\bf 619}, 193 (2005);
K.A. Bronnikov and M.S. Chernakova, gr-qc/0503025;
M.C.B. Abdalla, S. Nojiri and S.D. Odintsov, {\em Class. Quant. 
Grav.} {\bf 22}, L35 (2005).

\bibitem{Sotiriou} T.P. Sotiriou, gr-qc/0604028; gr-qc/0509029; gr-qc/0512017.

\bibitem{Palatini} 
D.N. Vollick, {\em Phys. Rev. D} {\bf 68}, 
063510 (2003); {\em Class. Quant. Grav.} {\bf 21}, 3813 (2004);
X.H. Meng and P. Wang, {\em Class. Quant. 
Grav.} {\bf 20}, 4949 (2004);  {\em Class. Quant. Grav.} {\bf 
21}, 951 (2004); {\em Phys. Lett. B} {\bf 584}, 1 (2004);
E.E. Flanagan, {\em Phys. Rev. Lett.} {\bf 
92}, 071101 (2004);
T. Koivisto, {\em Phys. Rev. D} {\bf 73}, 083517 (2006); 
gr-qc/0505128; 
T. Koivisto and H. Kurki-Suonio, {\em Class. Quant. Grav.} {\bf 
23}, 2355 (2006); 
P. Wang, G.M. Kremer, D.S.M. Alves, and X.H. Meng, 
{\em Gen. Rel. Grav.} {\bf 38}, 517 (2006);
G. Allemandi, M. Capone, S. Capozziello, and M. Francaviglia, 
{\em Gen. Rel. Grav.} {\bf 38}, 33 (2006);

\bibitem{LiberatiSotiriou} T.P. Sotiriou and S. Liberati, 
gr-qc/0604006;
N.J. Poplawski, {\em Class. Quant. Grav.} {\bf 23}, 2011 (2006); 
gr-qc/051107.

\bibitem{TeyssandierTourrenc} P. Teyssandier and P. Tourrenc, 
{\em J. Math. Phys.} {\bf 24}, 2793 (1983).

\bibitem{Wands} D. Wands, {\em Class. Quant. Grav.} {\bf 11}, 
269 (1994).

\bibitem{ChibaPLB03} T. Chiba, {\em Phys. Lett. B} {\bf 575}, 1 
(2005).

\bibitem{TorresVucetich} D.F. Torres and H. Vucetich, {\em Phys. Rev. D} 
{\bf 54}, 7373 (1996).

\bibitem{BransDicke} C.H. Brans and R.H. Dicke, {\em Phys. Rev.} 
{\bf 124}, 925 (1961).

\bibitem{ST} P.G. Bergmann, {\em Int. J. Theor. Phys.}  {\bf 1},  
 25 (1968);
R.V. Wagoner, {\em Phys. Rev. D} {\bf 1},  3209  (1970);
K. Nordvedt, {\em Astrophys. J.} {\bf 161},  1059 (1970).

\bibitem{mybook} V. Faraoni, {\em Cosmology in Scalar-Tensor 
Gravity} (Kluwer Academic, Dordrecht, 2004).

\bibitem{PPN} M.E. Soussa and R.P. Woodard, {\em Gen. Rel. 
Grav.} {\bf 36}, 855 (2004);  
R. Dick, {\em Gen. Rel. Grav.} {\bf 36}, 217 (2004);
A.E. Dominguez and D.E. Barraco, {\em Phys. Rev. D} {\bf 70}, 
043505 (2004);
D.A. Easson, {\em Int. J. Mod. Phys. A} {\bf 19}, 5343 (2004);
G.J. Olmo, {\em Phys. Rev. Lett.} {\bf 95}, 261102 (2005);
{\em Phys. Rev. D} {\bf 72}, 083505 (2005);
gr-qc/0505135; gr-qc/0505136;
I. Navarro and K. Van Acoleyen, {\em Phys. Lett. B} {\bf 622}, 1 
(2005);
G. Allemandi, M. Francaviglia, M.L. Ruggiero and A. Tartaglia, 
{\em Gen. Rel. Grav.} {\bf 37}, 1891 (2005);
J.A.R. Cembranos, {\em Phys. Rev. D} {\bf 73}, 064029 (2006);
S. Capozziello and A. Troisi, {\em Phys. Rev. D} {\bf 72}, 
044022 (2005);
T. Clifton and J.D. Barrow, {\em Phys. Rev. D} {\bf 72}, 103005 
(2005).

\bibitem{SotiriouPPN} T.P. Sotiriou, gr-qc/0507027.

\bibitem{footnote1} A 
cosmological constant $\Lambda$ can be added to the linear part 
of $f(R)$ but we omit it because modified gravity was 
introduced as an alternative to dark energy and the cosmological 
constant.

\bibitem{footnote2} This 
procedure has been considered for the specific scenario 
$f(R) =R-\mu^4/R $ \cite{ShaoCaiWangSuPLB}, where the limit 
$\mu\rightarrow 0$ is effectively replaced by an expansion in 
the small parameter $\mu\approx H_0 \approx 10^{-33}$~eV 
analogous to $\epsilon$. However, this scenario is not 
viable due to instabilities in the matter 
\cite{DolgovKawasaki} and the gravity \cite{FaraoniNadeau} 
sectors unless a positive power of $R$ is added to the 
Lagrangian.

\bibitem{DolgovKawasaki} A.D. Dolgov and M. Kawasaki, {\em Phys. 
Lett. B} {\bf 573}, 1 (2003).

\bibitem{FaraoniNadeau} V. Faraoni and S. Nadeau, {\em Phys. 
Rev. D} {\bf 72}, 124005 (2005).

\bibitem{Weinberg} S. Weinberg, {\em Gravitation and Cosmology} 
(Wiley, New York, 1972).

\bibitem{exactBDsolutions} 
T. Matsuda, {\em Progr. Theor. Phys.} {\bf 47},
738 (1972);
C. Romero and A. Barros, {\em Astrophys. Sp.
Sci.} {\bf 192}, 263 (1992);
preprint DF-CCEN-UFPb No. 9 (1992);
{\em Phys. Lett. A} {\bf 173}, 243 (1993);
{\em Gen. Rel. Grav.} {\bf 25}, 491 (1993);
F.M. Paiva, M. Reboucas, and M. MacCallum, {\em
Class. Quant. Grav.} {\bf 10}, 1165 (1993);
F.M. Paiva and C. Romero, {\em Gen. Rel. Grav.} {\bf 25}, 1305 
(1993);
M.A. Scheel, S.L. Shapiro, and S.A. Teukolsky, {\em Phys. Rev. D} 
{\bf 51}, 4236 (1995);
N. Banerjee and S. Sen, {\em Phys. Rev. D} {\bf
56}, 1334 (1997);
L.A. Anchordoqui, D.F. Torres, M.L. Trobo, and G.S. 
Birman, {\em Phys. Rev. D} {\bf 57}, 829 (1998).

\bibitem{VFBDlimit} V. Faraoni, {\em Phys. Lett. A} {\bf 245}, 
26 (1998); {\em Phys. Rev. D} {\bf 59}, 084021 (1999).

\bibitem{footnote3} Recent 
developments on this issue 
include the discovery of certain solutions corresponding to 
energy-momentum tensor with 
$T^{(m)}\neq 0$ which also fail to converge to the corresponding 
solutions of Einstein's theory when $\omega\rightarrow \infty$ 
\cite{southamericans,Chauvineau}, 
and even the conjecture that the failure to achieve the limit to 
GR when $\omega $ becomes large may be a generic feature of 
Brans-Dicke gravity \cite{Chauvineau,BhadraNandi}.

\bibitem{southamericans} L.A. Anchordoqui, S.P. Bergliaffa, M.L. 
Trobo, and G.S. Birman, {\em Mod. Phys. Lett. A} {\bf 14}, 1105 
(1999).

\bibitem{Chauvineau} B. Chauvineau, {\em Class. Quant. Grav.} 
{\bf 20}, 2617 (2003).

\bibitem{BhadraNandi} A. Bhadra and K.K. Nandi, {\em Phys. Rev. 
D} {\bf 64}, 087501 (2001).

\bibitem{BanerjeeSen} N. Banerjee and S. Sen, {\em Phys. Rev. D} 
{\bf 56}, 1334 (1997).

\bibitem{VFrapid} V. Faraoni, {\em Phys. Rev. D} {\bf  72}, 
061501(R) (2005).

\bibitem{footnote4} The study of physically different 
instabilities yields surprisingly similar stability conditions. 
See Refs.~\cite{NavarrovanAcoleyen, 
Cognolaetal, DolgovPelliccia, 
BarrowHervik, MullerSchmidtStarobinsky,Orfeu} 
for alternative studies of the stability of de Sitter space.

\bibitem{NavarrovanAcoleyen} I. Navarro and K. van Acoleyen, 
{\em J. Cosmol. Astropart. Phys.} {\bf 0603}, 008 (2006).

\bibitem{Cognolaetal} G. Cognola, E. Elizalde, S. Nojiri, S.D. 
Odintsov, and S. Zerbini, {\em J. Cosmol. Astropart. Phys.} {\bf 
02}, 010 (2005); {\em J. Phys. A} {\bf 39}, 6245 (2006).

\bibitem{DolgovPelliccia} A. Dolgov and D.N. Pelliccia, 
hep-th/0502197.

\bibitem{MullerSchmidtStarobinsky} V. M\"{u}ller, 
H.-J. Schmidt, and A.A. Starobinsky, {\em Phys. Lett. B} {\bf 
202}, 198 (1988); H.-J. Schmidt, {\em Class. Quant. Grav.} {\bf 
5}, 233 (1988); A. Battaglia Mayer and H.-J. Schmidt, {\em Class. 
Quant. Grav.} {\bf 10}, 244 (1993).

\bibitem{BarrowHervik} J.D. Barrow and S. Hervik, {\em Phys. 
Rev. D} {\bf 73}, 023007 (2006).

\bibitem{Orfeu} O. Bertolami, {\em Phys. Lett. B} {\bf 186}, 161 
(1987).

\bibitem{VFMJ} V. Faraoni and M.N. Jensen, {\em Class. Quant. 
Grav} {\bf 23}, 30005 (2006).

\bibitem{Bardeen} J.M. Bardeen, {\em Phys. Rev. D} {\bf 22},  
1882 (1980);
G.F.R. Ellis and M. Bruni, {\em Phys. Rev. D} {\bf 40}, 1804 
(1989); 
G.F.R. Ellis, J.C. Hwang and M. Bruni, {\em Phys. Rev. D} 
{\bf 40},  1819 (1989); 
G.F.R. Ellis, M. Bruni and J.C. Hwang, {\em Phys. Rev. D} 
{\bf 42}, 1035 (1990).


\bibitem{Hwang} J.C. Hwang, {\em Class. Quant. Grav.}  {\bf 
7},  1613 (1990);
{\bf 14},  1981 (1997); 3327; {\bf 15},  1401 (1998);  1387; 
{\em Phys. Rev. D} {\bf 42},  2601 (1990); {\bf 53},  762 (1996); 
J.C. Hwang and H. Noh, {\em Phys. Rev. D} {\bf 54}, 1460 (1996).


\bibitem{Bertotti} B. Bertotti, L. Iess, and P. Tortora, {\em 
Nature} {\bf 425}, 374 (2003).

\bibitem{MGviable} S. Capozziello, A. Stabile,  and A. Troisi, 
gr-qc/0603071.

\bibitem{ShaoCaiWangSuPLB} C.-G. Shao, R.-G. Cai, B. Wang, and 
R.-K. Su, {\em Phys. Lett. B} {\bf 633}, 164 (2006).

\bibitem{instabilities} A. N\'{u}nez and S. Solganik, {\em Phys. 
Lett. B} {\bf 608}, 189 (2005); hep-th/0403159;
T. Chiba, {\em J. Cosmol. Astropart. Phys.} {\bf 0505}, 008 
(2005);
P. Wang, {\em Phys. Rev. D} {\bf 72}, 024030 (2005);
A. De Felice, M. Hindmarsh, and M. Trodden, astro-ph/0604154.

\end{thebibliography}

\end{document}